\documentclass{rnaastex}

\newcommand{\GBTSEFDC}{24.8}
\newcommand{\GBTSEFDS}{55.6}
\newcommand{\KMSEFD}{2536.6}

\definecolor{black}{rgb}{0,0,0}
\definecolor{red}{rgb}{1.0,0,0}

\newcommand{\UCB}{Department of Astronomy, 501 Campbell Hall, University of California Berkeley, Berkeley CA 94720}
\newcommand{\SSL}{Space Sciences Laboratory, University of California, Berkeley, Berkeley CA 94720}
\newcommand{\SWIN}{Centre for Astrophysics \& Supercomputing, Swinburne University of Technology, PO Box 218, Hawthorn, VIC 3122, Australia}
\newcommand{\GBT}{Green Bank Observatory, PO Box 2 Green Bank, West Virginia, 24944, USA}
\newcommand{\NIJ}{Department of Astrophysics/IMAPP,Radboud University, Nijmegen, Netherlands}
\newcommand{\SETI}{SETI Institute, Mountain View, California}
\newcommand{\CALTECH}{California Institute of Technology, 1200 E California Blvd MC 249-17, Pasadena, CA 91125, USA}
\newcommand{\PEKING}{Department of Astronomy \& Kavli Institute for Astronomy and Astrophysics, Peking University}
\newcommand{\YUNNAN}{Yunnan Astronomical Observatory, Chinese Academy of Sciences, Kunming 650011, China}
\newcommand{\XINJIANG}{Xinjiang Astronomical Observatory, Chinese Academy of Sciences,Urumqi 830011, China}
\begin{document}

\title{No bursts detected from FRB121102 in two 5-hour observing campaigns with the Robert C. Byrd Green Bank Telescope}

\correspondingauthor{Danny C. Price}
\email{dancpr@berkeley.edu}


\author[0000-0003-2783-1608]{Danny C.\ Price}
\affiliation{\UCB}
\affiliation{\SWIN}

\author[0000-0002-8604-106X]{Vishal Gajjar}
\affiliation{\SSL}

\author{Lee Rosenthal}
\affiliation{\CALTECH}

\author{Gregg Hallinan}
\affiliation{\CALTECH}

\author[0000-0003-4823-129X]{Steve Croft}
\affiliation{\UCB}

\author[0000-0003-3197-2294]{David DeBoer}
\affiliation{\UCB}

\author{Greg Hellbourg}
\affiliation{\UCB}

\author[0000-0002-0531-1073]{Howard Isaacson}
\affiliation{\UCB}

\author{Matt Lebofsky}
\affiliation{\UCB}

\author{Ryan Lynch}
\affiliation{\GBT}

\author{David H.\ E.\ MacMahon}
\affiliation{\UCB}

\author{Yunpeng Men}
\affiliation{\PEKING}

\author{Yonghua Xu}
\affiliation{\YUNNAN}
\author{Zhiyong Liu}
\affiliation{\XINJIANG}
\author{Kejia Lee}
\affiliation{\PEKING}
\author[0000-0003-2828-7720]{Andrew Siemion}
\affiliation{\UCB}
\affiliation{\NIJ}
\affiliation{\SETI}




\section{Introduction}
\label{sec:intro}

More than 10 years after the discovery of Fast Radio Bursts (FRBs; see \citealt{Lorimer:2007}), an understanding of their origins remains elusive. Of the 29 reported FRBs\footnote{\url{http://www.frbcat.org}}, only one of these, FRB 121102, has been shown to repeat \citep{Spitler:2016}. This has allowed numerous follow-up campaigns, resulting in an unambiguous localization to its host galaxy \citep{Chatterjee:2017,Marcote:2017}. 
Multi-wavelength campaigns to characterize and monitor its spectral index, burst rate, polarization, and spectro-temporal variations \cite[e.g.][]{Scholz:2016, Gajjar:2017, Michilli:2018} are ongoing. 

Here, we report non-detection of radio bursts from FRB 121102 during two 5-hour observation sessions on the Robert C. Byrd 100-m Green Bank Telescope in West Virginia, US (GBT) on December 11, 2017, and January 12, 2018. 
In addition, we report non-detection during an abutting 10-hour observation with the Kunming 40-m telescope in China (KM40), which commenced UTC 10:00 January 12, 2018. 
These are among the longest published contiguous observations of FRB 121102, and support the notion that FRB 121102 bursts are episodic. 

These observations were part of a simultaneous optical and radio monitoring campaign with the the Caltech HIgh-speed Multi-color CamERA \citep[CHIMERA,][]{Harding:2016} instrument on the Hale 5.1-m telescope. The data analysis of CHIMERA data is ongoing and will be published elsewhere. 


\section{Observations}
\label{sec:obs}

We observed FRB 121102 for two 5-hour sessions with the GBT, commencing UTC 2017-12-11T03:00 and 2018-01-12T02:30 using the Breakthrough Listen digital backend \citep{Macmahon:2017} to record baseband data across the nominal bands of the receivers.
On  11th December, 2017, we observed using the 4.0--8.0~GHz receiver (SEFD\footnote{SEFD: System Equivalent Flux Density} $\sim$10~Jy); at the start of observations we ran the GBT autopeak focus routine to calibrate the active surface. On 12th January, 2018, we observed using the 1.6--2.6~GHz receiver (SEFD $\sim$10~Jy); no autopeak calibration is required at these frequencies. During both sessions, we observed 3C161 and PSR B0525+21 for flux and polarization calibration. Observations of FRB 121102 were conducted in 30-minute segments (Table 1). 

From UTC 2018-01-12T10:00 onwards, the source became visible to the KM40 telescope, using which a 10-hour observation was carried out with the newly-installed 4.7--5.2 GHz receiver (SEFD $\sim$256~Jy).

\section{Results and Discussion}

At the GBT, baseband data were reduced to form high time resolution (300 $\mu$s) Stokes-I dynamic spectra (183~kHz frequency resolution) using the Breakthrough Listen GPU-accelerated spectroscopy suite. The reduced products were searched for dispersed pulses consistent with the known 557 pc cm$^{-3}$ dispersion measure of FRB 121102, using the \textsc{Heimdall} software package \citep{Barsdell:2012}. 
At the KM40, Stokes-I dynamic spectra (64~$\mu$s,  1~MHz) were recorded and searched in real-time using the \textsc{Bear} software package (details forthcoming).

No bursts were detected during either session. In contrast, 15 bursts were detected within 30 minutes in previous GBT observations over 4.0--8.0~GHz using the same procedure \citep[][Gajjar et. al., in prep]{Gajjar:2017}. Taken together, these observations support models that predict episodic emission \citep{Scholz:2016}. We have published these non-detections here foremostly so that a better statistical model can be formed by combination with burst statistics from other observing campaigns.

\acknowledgments

Breakthrough Listen is managed by the Breakthrough Initiatives, sponsored by the \href{http://breakthroughinitiatives.org}{Breakthrough Prize Foundation}. Part of this research was carried out at the Jet Propulsion Laboratory, California Institute of Technology, under a contract with the National Aeronautics and Space Administration. VG would like to acknowledge NSF grant 1407804 and the Marilyn and Watson Alberts SETI Chair funds. KJL and YPM were supported by NSFC U15311243, XDB23010200 and Max-Planck partner group with MPIfR. LZY was supported by foundation for Key laboratory of Xinjiang Uygur Autonomous Region (2015KL012).

\bibliographystyle{aasjournal}
\bibliography{references}

\begin{thebibliography}{}
\expandafter\ifx\csname natexlab\endcsname\relax\def\natexlab#1{#1}\fi
\providecommand{\url}[1]{\href{#1}{#1}}

\bibitem[{{Barsdell} {et~al.}(2012){Barsdell}, {Bailes}, {Barnes}, \&
  {Fluke}}]{Barsdell:2012}
{Barsdell}, B.~R., {Bailes}, M., {Barnes}, D.~G., \& {Fluke}, C.~J. 2012,
  MNRAS, 422, 379

\bibitem[{Chatterjee {et~al.}(2017)Chatterjee, Law, Wharton, Burke-Spolaor,
  Hessels, Bower, Cordes, Tendulkar, Bassa, Demorest, Butler, Seymour, Scholz,
  Abruzzo, Bogdanov, Kaspi, Keimpema, Lazio, Marcote, McLaughlin, Paragi,
  Ransom, Rupen, Spitler, \& van Langevelde}]{Chatterjee:2017}
Chatterjee, S., Law, C.~J., Wharton, R.~S., {et~al.} 2017, Nature, 541, 58

\bibitem[{{Gajjar} {et~al.}(2017){Gajjar}, {Siemion}, {MacMahon}, {Croft},
  {Hellbourg}, {Isaacson}, {Enriquez}, {Price}, {Lebofsky}, {DeBoer},
  {Werthimer}, {Hickish}, {Brinkman}, {Chatterjee}, \& {Ransom}}]{Gajjar:2017}
{Gajjar}, V., {Siemion}, A.~P.~V., {MacMahon}, D.~H.~E., {et~al.} 2017, The
  Astronomer's Telegram, 10675

\bibitem[{{Harding} {et~al.}(2016){Harding}, {Hallinan}, {Milburn}, {Gardner},
  {Konidaris}, {Singh}, {Shao}, {Sandhu}, {Kyne}, \&
  {Schlichting}}]{Harding:2016}
{Harding}, L.~K., {Hallinan}, G., {Milburn}, J., {et~al.} 2016, MNRAS, 457,
  3036

\bibitem[{{Lorimer} {et~al.}(2007){Lorimer}, {Bailes}, {McLaughlin},
  {Narkevic}, \& {Crawford}}]{Lorimer:2007}
{Lorimer}, D.~R., {Bailes}, M., {McLaughlin}, M.~A., {Narkevic}, D.~J., \&
  {Crawford}, F. 2007, Science, 318, 777

\bibitem[{{MacMahon} {et~al.}(2017){MacMahon}, {Price}, {Lebofsky}, {Siemion},
  {Croft}, {DeBoer}, {Enriquez}, {Gajjar}, {Hellbourg}, {Isaacson},
  {Werthimer}, {Abdurashidova}, {Bloss}, {Creager}, {Ford}, {Lynch},
  {Maddalena}, {McCullough}, {Ray}, {Whitehead}, \& {Woody}}]{Macmahon:2017}
{MacMahon}, D.~H.~E., {Price}, D.~C., {Lebofsky}, M., {et~al.} 2017, ArXiv
  e-prints, arXiv:1707.06024

\bibitem[{Marcote {et~al.}(2017)Marcote, Paragi, Hessels, Keimpema, van
  Langevelde, Huang, Bassa, Bogdanov, Bower, Burke-Spolaor, Butler, Campbell,
  Chatterjee, Cordes, Demorest, Garrett, Ghosh, Kaspi, Law, Lazio, McLaughlin,
  Ransom, Salter, Scholz, Seymour, Siemion, Spitler, Tendulkar, \&
  Wharton}]{Marcote:2017}
Marcote, B., Paragi, Z., Hessels, J. W.~T., {et~al.} 2017, ApJL, 834, L8

\bibitem[{Michilli {et~al.}(2018)Michilli, Seymour, Hessels, Spitler, Gajjar,
  Archibald, Bower, Chatterjee, Cordes, Gourdji, Heald, Kaspi, Law, Sobey,
  Adams, Bassa, Bogdanov, Brinkman, Demorest, Fernandez, Hellbourg, Lazio,
  Lynch, Maddox, Marcote, McLaughlin, Paragi, Ransom, Scholz, Siemion,
  Tendulkar, Van~Rooy, Wharton, \& Whitlow}]{Michilli:2018}
Michilli, D., Seymour, A., Hessels, J. W.~T., {et~al.} 2018, Nature, 553, 182

\bibitem[{Scholz {et~al.}(2016)Scholz, Spitler, Hessels, Chatterjee, Cordes,
  Kaspi, Wharton, Bassa, Bogdanov, Camilo, Crawford, Deneva, van Leeuwen,
  Lynch, Madsen, McLaughlin, Mickaliger, Parent, Patel, Ransom, Seymour,
  Stairs, Stappers, \& Tendulkar}]{Scholz:2016}
Scholz, P., Spitler, L.~G., Hessels, J. W.~T., {et~al.} 2016, The Astrophysical
  Journal, 833, 177

\bibitem[{Spitler {et~al.}(2016)Spitler, Scholz, Hessels, Bogdanov, Brazier,
  Camilo, Chatterjee, Cordes, Crawford, Deneva, Ferdman, {Freire, P. C. C.},
  Kaspi, Lazarus, Lynch, Madsen, McLaughlin, Patel, Ransom, Seymour, Stairs,
  Stappers, van Leeuwen, \& Zhu}]{Spitler:2016}
Spitler, L.~G., Scholz, P., Hessels, J. W.~T., {et~al.} 2016, Nature, 531, 202

\end{thebibliography}

\begin{table*}
\centering
\caption{Details of FRB 121102 observing sessions on GBT and KM40. No radio bursts were detected during these periods.}
\begin{tabular}{rcccccc}
\hline
Telescope & Scan ID &  Frequency band & Flux limit$^{\dag}$  & \multicolumn{2}{c}{Observation start date} & Duration \\
 & & (GHz) & (mJy) & (UTC) & (MJD) & (min) \\
\hline
\hline
GBT & 20171211-1  & 4.0--8.0  & \GBTSEFDC & 2017-12-11T03:46:40.000 & 58098.1574074074 & 30  \\
GBT & 20171211-2  & 4.0--8.0  & \GBTSEFDC & 2017-12-11T03:47:50.000 & 58098.1582175925 & 30  \\
GBT & 20171211-3  & 4.0--8.0  & \GBTSEFDC & 2017-12-11T04:18:12.000 & 58098.1793055555 & 30  \\
GBT & 20171211-4  & 4.0--8.0  & \GBTSEFDC & 2017-12-11T04:48:22.000 & 58098.2002546296 & 30  \\
GBT & 20171211-5  & 4.0--8.0  & \GBTSEFDC & 2017-12-11T05:44:14.000 & 58098.2390509259 & 30  \\
GBT & 20171211-6  & 4.0--8.0  & \GBTSEFDC & 2017-12-11T06:14:50.000 & 58098.2603009259 & 30  \\
GBT & 20171211-7  & 4.0--8.0  & \GBTSEFDC & 2017-12-11T06:45:01.000 & 58098.2812615740 & 30  \\
GBT & 20171211-8  & 4.0--8.0  & \GBTSEFDC & 2017-12-11T07:15:12.000 & 58098.3022222222 & 30  \\
GBT & 20171211-9  & 4.0--8.0  & \GBTSEFDC & 2017-12-11T07:45:22.000 & 58098.3231712962 & 14  \\
\hline
GBT & 20180112-1  & 1.6--2.6  & \GBTSEFDS & 2018-01-12T02:44:37.000 & 58130.1143171296 & 30  \\
GBT & 20180112-2  & 1.6--2.6  & \GBTSEFDS & 2018-01-12T03:14:46.000 & 58130.1352546296 & 30  \\
GBT & 20180112-3  & 1.6--2.6  & \GBTSEFDS & 2018-01-12T03:44:55.000 & 58130.1561921296 & 30  \\
GBT & 20180112-4  & 1.6--2.6  & \GBTSEFDS & 2018-01-12T04:15:04.000 & 58130.1771296296 & 30  \\
GBT & 20180112-5  & 1.6--2.6  & \GBTSEFDS & 2018-01-12T04:45:13.000 & 58130.1980671296 & 30  \\
GBT & 20180112-6  & 1.6--2.6  & \GBTSEFDS & 2018-01-12T05:15:22.000 & 58130.2190046296 & 30  \\
GBT & 20180112-7  & 1.6--2.6  & \GBTSEFDS & 2018-01-12T05:45:31.000 & 58130.2399421296 & 30  \\
GBT & 20180112-8  & 1.6--2.6  & \GBTSEFDS & 2018-01-12T06:15:40.000 & 58130.2608796296 & 30  \\
GBT & 20180112-9  & 1.6--2.6  & \GBTSEFDS & 2018-01-12T07:03:04.000 & 58130.2937962962 & 28  \\
\hline
KM40 & 20180112-10  & 4.7--5.2   & \KMSEFD  & 2018-01-12T09:59:53.862 & 58130.4165956206  & 600 \\
\hline
\multicolumn{5}{l}{$^{\dag}$ assuming pulse width of 1~ms, detection SNR threshold of 7$\sigma$}\\
\end{tabular}
\label{table:obs_table}
\end{table*}

\end{document}